# Two-Level Systems and Boson Peak Remain Stable in 110-Million-Year-Old Amber Glass

Tomás Pérez-Castañeda,[1] Rafael J. Jiménez-Riobóo,[2] and Miguel A. Ramos[1][‡]

[1]Laboratorio de Bajas Temperaturas, Departamento de Física de la Materia Condensada, Condensed Matter Physics Center (IFIMAC) and Instituto Nicolás Cabrera, Universidad Autónoma de Madrid, E-28049 Madrid, Spain.
[2]Instituto de Ciencia de Materiales de Madrid, Consejo Superior de Investigaciones Científicas (ICMM-CSIC), E-28049 Madrid, Spain.

**Abstract** The two most prominent and ubiquitous features of glasses at low temperatures, namely the presence of tunneling two-level systems and the so-called *boson peak* in the reduced vibrational density of states, are shown to persist essentially unchanged in highly stabilized glasses, contrary to what was usually envisaged. Specifically, we have measured the specific heat of 110 million-year-old amber samples from El Soplao (Spain), both at very low temperatures and around the glass transition $T_g$. In particular, the amount of two-level systems, assessed at the lowest temperatures, was surprisingly found to be exactly the same for the pristine *hyperaged* amber as for the subsequently partially- and fully-rejuvenated samples.

PACS numbers: 65.60.+a, 63.50.Lm, 64.70.P-, 81.40.Cd

Amber has been appreciated by mankind for its color and natural beauty since the Neolithic era. It also has a well-known paleontological significance, being a unique preservational system where ancient bio-inclusions of animal and plant materials were stuck in the viscous resin and then fossilized for million years [1, 2]. Specifically, amber is a fossilized tree resin produced from the exudates of conifers or angiosperms which has undergone a maturation process over geological time, where progressive polymerization takes place as well as evaporation of volatile components, isomerization reactions, crosslinking and cyclization. As a result of all this, resins have been able to fossilize, after long periods of time that can exceed one hundred million years, into extremely stable materials: amber glasses. Several amber bearing deposits around the world with different types or chemical compositions of amber have proved invaluable for paleontology in the reconstruction of ecosystems and prehistoric life [1, 2].

For physics and chemistry research, amber is a unique example of a glass that has been aging for a very long time below its glass transition temperature, thus reaching a state which is not accessible under normal experimental conditions. From a chemical point of view, amber is a macromolecular solid resulting of free radical polymerization. From a physical point of view, it is an amorphous solid or glass which has experienced an extreme thermodynamic stabilization process (*hyperaging*). As a matter of fact, amber, as many other natural or synthetic polymers, is a glass produced by *chemical* vitrification [3], whereas standard chemically stable glass-forming liquids become glasses by *physical* vitrification, i.e. by reducing temperature or increasing pressure. Nevertheless, both kinds of glass have been shown to exhibit essentially the same kinetic and thermodynamic properties characterizing the glassy behavior [3, 4].

Our understanding of the particularly rich phenomenology of glasses and supercooled liquids continues to be a major unsolved scientific challenge [5–10]. Whether the glass transition itself is only a purely kinetic event or the manifestation of an underlying thermodynamic transition (occurring, for instance, at the Kauzmann [11] temperature $T_K$ where the extrapolated entropy of the glass-forming liquid would equal that of the crystal state, if much more slowly cooled) is still under debate [6, 7]. A very useful framework for interpreting the complex phenomenology of glasses and supercooled liquids is provided [12] by the potential-energy landscape (PEL). As depicted in Fig. 1, the PEL is a topographic view of the ($3N+1$) potential-energy hypersurface of any glass-forming substance of $N$ particles, schematically projected on two dimensions for convenience, which has many local minima and saddle points for thermal energies below that of the melting point for the stable crystalline state (absolute minimum).

Thus, amber provides a unique benchmark to study the properties of a glass very close to its "ideal-glass" limit (as if it could be brought in thermodynamic metastable equilibrium down to $T_K$), to be followed by a comparison with the canonical glass obtained by subsequently erasing the thermal history of the amber glass (rejuvenation). One could investigate, for example, which characteristic properties of glasses are robust and inherent to the non-crystalline state, and which others are dependent on the degree of frozen-in disorder at the glass transition –and so might disappear on aging.

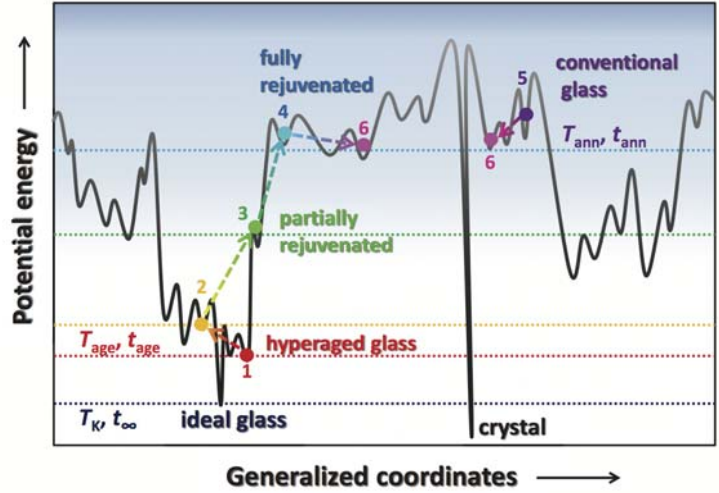

FIG. 1. Schematic potential energy landscape (PEL) for supercooled liquids and glasses, including the hypothetical absolute minimum for a crystal state. An ideal glass would be obtained after an infinitely long aging at the Kauzmann temperature $T_K$. The likely level for the studied hyperaging process is indicated by $T_{age}$, $t_{age}$. The successive states obtained with the gradual isothermal annealings (rejuvenation) at some $T_{ann}$, $t_{ann}$ applied to the pristine sample of amber, are illustrated with the path described by the circles and the arrows. The locations labeled 1−6 in this PEL serve to identify the different samples throughout the Letter.

Specifically, our work has been aimed at studying the universal *anomalous* properties exhibited by glasses at low temperatures [8–10], using amber as a model system. In brief, glasses and other amorphous solids systematically exhibit at low temperature a specific heat $C_p$ much larger and a thermal conductivity $\kappa$ orders of magnitude lower, respectively, than the corresponding values found in their crystalline counterparts [13]. Furthermore, $C_p$ was found [8, 13] to depend quasilinearly ($C_p \propto T^{1+\delta}$) and $\kappa$ almost quadratically ($\kappa \propto T^{2-\delta}$) on temperature $T$, in clear contrast to the cubic dependences observed in crystals for both properties, well understood in terms of Debye's theory of lattice vibrations. These and related acoustic and dielectric properties of amorphous solids at low temperatures [8] were soon well accounted for [14, 15] by the successful Tunneling Model (TM), though some open questions remain unsolved [16].

On the other hand, the thermal behavior of glasses above 1 K and their corresponding low-frequency vibrational properties around 1 THz, are much more poorly understood. This frequency range is indeed dominated by another universal and much disputed feature of glasses: the so-called "boson peak" [8, 10] arising from a noteworthy excess in the vibrational density of states (VDOS) over that predicted by Debye's theory $g(\omega) \propto \omega^2$. Such an excess in the low-frequency VDOS appears as a broad peak in $g(\omega)/\omega^2$, which produces the broad maximum in $C_p/T^3$ observed in most glasses at a few K.

**TABLE I.** Calorimetric (devitrification) glass-transition temperatures $T_g^*$ and fictive temperatures $T_f$ obtained after the different thermal histories applied to the studied samples. The expected location in the potential-energy landscape (PEL) of Fig. 1 is indicated in the second column and serves as label for the samples. The last column displays the relative decrease of the fictive temperature $T_f$ in relation to the standard glass-transition temperature of this Spanish amber, $T_g$= 423 K.

| **Thermal history** | # PEL | $T_g^*$ (K) | $T_f$ (K) | $(T_f - T_g)/T_g$ |
|---|---|---|---|---|
| Pristine (hyperaged) | 1 | 438 | 384 | −9.2% |
| 3 h @ 393 K | 2 | 438 | 385 | −9.0% |
| 2 h @ 423 K | 3 | 436 | 391 | −7.6% |
| 1.5 h @ 433 K | 4a | 423 | 413 | −2.4% |
| 1 h @ 443 K | 4b | 423 | 415 | −1.9% |
| Rejuvenated (>460 K) | 4c | 423 | 416 | −1.7% |
| Quenched | 5 | 423 | 417 | −1.4% |
| Rejuvenated & annealed 2 h @ 423 K | 6 | 424 | 416 | −1.7% |

In this Letter, we have measured the specific heat of several amber samples from El Soplao, in the low-temperature range 0.07K<$T$<30K, through their gradual rejuvenation and corresponding characterization, including elasto-optic measurements below room temperature and temperature-modulated differential scanning calorimetry (TM-DSC) around the glass transition. The details about the experimental methods and calculations are included in the Supplemental Material [17].

Specific-heat curves around the glass transition in different states of amber, ranging from a pristine sample to one quenched from the liquid, are plotted in Fig. 2(a). Several different isothermal annealing processes were applied to pristine amber samples, trying to follow the route depicted in the PEL of Fig. 1. As expected, these annealing treatments below $T_g$ destabilize and increase the internal energy of the previously hyperaged glass, oppositely to usual annealing processes for conventional glasses. In Fig. 2(b) we present the corresponding enthalpy curves, obtained by integration of the former curves through $\Delta H = \int_T^{T_0} C_P(T)dT$, with $T_0$ being a reference temperature in the liquid state.

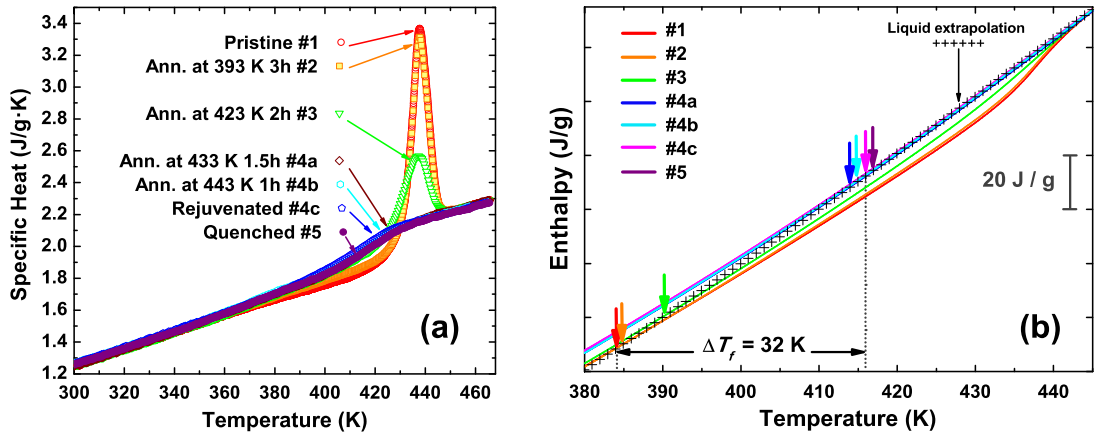

FIG. 2. (a) Specific heat of amber from *El Soplao* at different states, measured by TM-DSC with heating rates of 1 K/min and modulating amplitude ± 0.5 K/min every 80 s. Only the first upscan (see the Supplemental Material [17]) for each sample is presented. The devitrification temperature of the pristine sample is located at $T_g^*$=438 K, and then decreases with decreasing stability (rejuvenation). The aging signal is clearly seen as a huge endothermic peak at the glass transition, which is maximal in the case of the pristine sample 1. Rejuvenation of amber was done stepwise (Table I) by performing different isothermal treatments to the pristine sample near the glass transition. (b) Corresponding enthalpy curves from the specific-heat ones in (a), running from lower to upper solid curves when going from sample 1 to sample 5. The fictive temperature $T_f$ for each sample (indicated by the arrows, going from left to right) is obtained as the intersection of the extrapolated liquid curve (+) and the glass curve extrapolated from temperatures well below its $T_g^*$ (see the Supplemental Material [17]).

As can be seen in Fig. 2(a), a huge endothermic peak is observed for the pristine amber (sample 1) at the calorimetric devitrification temperature $T_g^*$ = 438 K (determined by the inflection point of the reversing $C_p$ jump [17]). $T_g^*$ is well above the genuine glass transition temperature $T_g$ = 423 K obtained for the rejuvenated sample 4c, or alternatively from the second or third heating runs for any sample [17], when the cooling and heating rates are canonically the same. This unusual *increase* of the calorimetric $T_g^*$ for the stabilized amber compared to the canonical glass, has been ascribed to a high kinetic stability in the related case of ultrastable thin films of organic glasses [18], indicating that much higher temperatures are needed to dislodge the molecules from their glassy configurations.

The fictive temperature $T_f$ is defined [19] as the temperature at which the nonequilibrium (glass) state and its equilibrium (supercooled liquid) state would have the same structure and properties, in particular enthalpy. As depicted in Fig. 2(b), we have determined $T_f$ as the intersection point (marked by the arrows) between the extrapolated enthalpy of the liquid and that of the corresponding glass at temperatures well below its $T_g^*$. Obtained data are given in Table I. The observed extraordinary *decrease* $\Delta T_f$ = 32 K (thermodynamic stability) for the pristine amber compared to the rejuvenated glass is similar or even superior to the effects seen in some ultrastable thin films of organic glasses [18, 20, 21]. Such a reduction over 9% of the fictive temperature with respect to the substance glass-transition is thus the consequence of extremely prolonged sub-sub-$T_g$ structural relaxations [22]. Notice that $T_g^* \neq T_f$ even for the canonical rejuvenated glass, what is mainly due to the specific calorimetric method employed to determine $T_g^*$ [17].

Following the theoretical route upwards through the PEL (Fig. 1), we experimentally see (Fig. 2) that applying an isothermal annealing for three hours at 393 K (sample 2), well below the glass transition $T_g$=423 K, only produces a very slight change in the pristine sample. When we further increase the annealing temperature, approaching it to $T_g$, both the endothermic $C_p$ peak and the enthalpy variation gradually decrease (sample 3, partial rejuvenation) until they completely disappear already for the thermal annealing at 433 K (sample 4a). Calorimetric curves after isothermal annealings (samples 4a and 4b) are indistinguishable from that for the fully-rejuvenated sample (sample 4c), obtained after heating it up to 470 K, well in the liquid state. Moreover, quenching the liquid at 50 K/min (sample 5) makes no significant difference from the conventional glass in the specific-heat and enthalpy curves. We have also tried to re-stabilize a rejuvenated amber glass, by annealing it for two hours at 423 K (sample 6), but it again produced almost negligible effects on both the thermodynamic and kinetic stability of the glass (Table I).

Finally, our main aim was to study the influence of the above-confirmed dramatic stabilization of the hyperaged glass on its low-temperature properties. We present in Fig. 3 our specific-heat measurements for pristine amber (sample 1), a partially rejuvenated sample (sample 3) and the fully rejuvenated one (sample 4c). Fig. 3(a) is a log-log plot at the lowest temperatures, which makes more clearly visible that the two-level-systems (TLS)-dominated low-temperature specific heat, remains invariable within experimental error. This is the main result of our work. On the other hand, above 1 K the specific heat moderately increases with rejuvenation around the "boson peak" in

$C_\mathrm{p}/T^3$ (occurring at 3.4±0.1 K in all cases, see Fig. 3(b)), following the same trend as the elastic Debye coefficient obtained from Brillouin-scattering and mass-density measurements (see Table II and the Supplemental Material [17]). As can be seen there, the aging process in amber has produced a densification of around 2%. The combination of mass-density and sound-velocity variations translates into a 9% lower Debye coefficient for the hyperaged (pristine) amber than for the rejuvenated (canonical) glass.

**TABLE II.** Measured mass density at room temperature $\rho_\mathrm{RT}$ and zero-temperature extrapolated $\rho(0)$, longitudinal $v_\mathrm{L}(0)$ and transverse $v_\mathrm{T}(0)$ sound velocity, average Debye velocity in the zero-temperature limit $v_\mathrm{D}$ and correspondingly calculated cubic Debye coefficient $c_\mathrm{D}$ for the specific heat [17], and height of the $C_\mathrm{p}/T^3$ boson peak.

| SAMPLE | $\rho_\mathrm{RT}$ (kg/m³) | $\rho(0)$ (kg/m³) | $v_\mathrm{L}(0)$ (m/s) | $v_\mathrm{T}(0)$ (m/s) | $v_\mathrm{D}$ (m/s) | $c_\mathrm{D}$ (µJ g⁻¹ K⁻⁴) | $(C_\mathrm{P}/T^3)_\mathrm{BP}$ (µJ g⁻¹ K⁻⁴) |
|---|---|---|---|---|---|---|---|
| **Pristine #1** | 1045 | 1055 | 3175 | 1635 | 1831 | **18.9** | 51.9 |
| **Annealed #3** | 1038 | 1049 | 3160 | 1625 | 1820 | **19.3** | 58.1 |
| **Rejuvenated #4c** | 1024 | 1035 | 3115 | 1596 | 1788 | **20.7** | 66.4 |

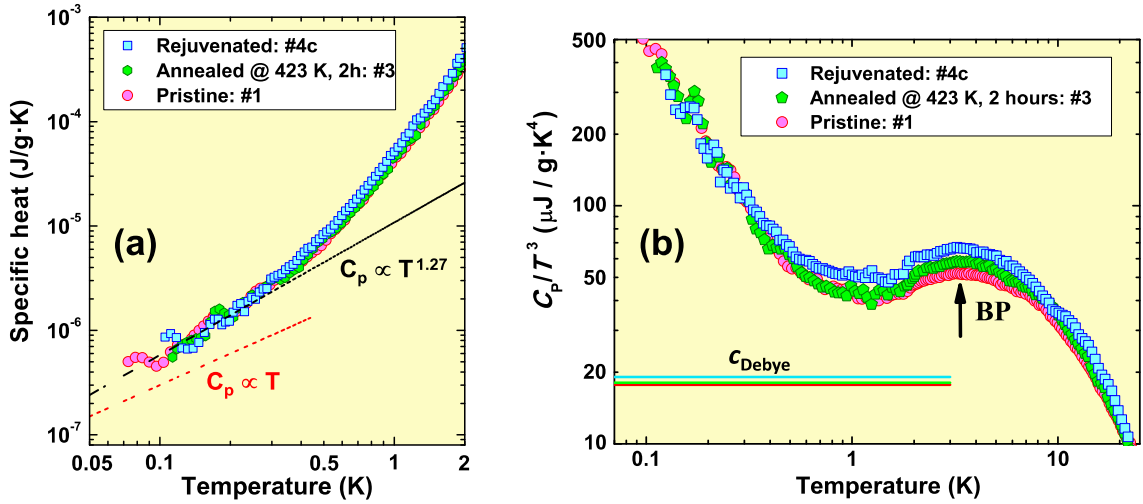

FIG. 3. (a) Comparison of specific-heat data for three Spanish amber samples (pristine, partially rejuvenated and fully rejuvenated) at very low temperatures, 0.05 K–2 K. The upper dashed line shows the best quasilinear fit to the experimental data below 0.4 K given by $C_\mathrm{p} \propto T^{1.27}$, hence faster than the simple linear dependence indicated by the lower dashed line. (b) $C_\mathrm{p}/T^3$ plot of the same data in (a), displayed in a wider temperature range. The height of the boson peak is observed to further increase with rejuvenation, though the values of both the minimum and the maximum of $C_\mathrm{p}/T^3$ remain constant: $T_\mathrm{min}$ = 1.2±0.1 K and $T_\mathrm{max}$ = 3.4±0.1 K, respectively. The corresponding Debye levels determined from the sound velocity and density data (Table II) are indicated by solid lines, and exhibit the same trend as the boson-peak height (both sets follow the same order as the symbols in the legend).

A strong $C_p/T^3$ peak was already observed by us [23] in pristine (20 million years old) Dominican amber, though apparent residual curing or repolymerization, occurring around the glass transition temperature when rejuvenating those amber samples, hindered a reliable quantitative investigation. Also stress-relaxation experiments in glass and supercooled-liquid states of Dominican amber have been recently reported [24].

Whether or not the low-temperature universal "anomalies" of glasses (i.e. TLS and boson peak) could be eventually suppressed by much stronger and longer annealing or aging processes, and hence whether they are or not intrinsic properties of the glass state, has been a long-standing question. During the last forty years, different experiments [25–30] have been reported about the possible influence of the thermal history on these properties, with contradictory conclusions. Many of them were focused on the paradigmatic vitreous silica. In brief, a modest decrease of the low-temperature specific heat with annealing was usually found [26], both at the lowest temperatures and around the $C_p/T^3$ maximum. However, the important role played by the different amounts of water content, as well as the lack of measured elasto-optic data with thermal treatment and hence of its variable contribution to the total heat capacity, makes their conclusions unclear. The same applies to early low-temperature specific-heat measurements in glassy glycerol [25], where the height of the $C_p/T^3$ maximum at 9 K was found to be a 7% lower in a slowly cooled glass than in a quenched glass. Also, a decrease with annealing of both the $C_p/T^3$ maximum and the $g(\omega)/\omega^2$ boson peak inferred from Raman scattering was reported in $As_2S_3$ glasses [27]. Nonetheless, later Raman and inelastic neutron scattering experiments [28] in the same glasses suggested that the observed changes in the total VDOS were caused by changes of sound velocity and density as a result of quenching. A similar conclusion was reached in dry $B_2O_3$ glasses [29], where very different $C_p/T^3$ peaks for different thermal treatments were found to merge into a single curve after their corresponding Debye levels had been subtracted. Interestingly, careful inelastic neutron scattering experiments in polybutadiene [30] showed a clear absence of annealing effect in its VDOS (boson peak), though a slight decrease in the boson peak was reported [31] after physical aging in another polymer, PMMA.

Indeed, some authors have tried to correlate the boson peak feature in glasses [32], and even in crystals [33], with transformations of the elastic continuum only. In the case of amber, however, such a Debye-scaling rule is not hold quantitatively. The height of the $C_p/T^3$ boson peak in the hyperaged amber has decreased a 22% from the standard rejuvenated glass, whereas a Debye scaling [32] ($\propto \omega_D^{-3}$, with $\omega_D$ being the Debye frequency) would predict only a 7.4% reduction. A similar relation was proposed by Shintani and Tanaka [34]. From numerical simulations in 2D glass-forming systems, they suggested that the boson peak height should scale with the inverse of the shear modulus. From the data in Table II, one would expect a boson-peak reduction < 7%, again well below the experimentally observed 22%.

Our experiments on hyperaged glasses of amber, far away from laboratory time-scale annealing or quenching processes, undoubtedly demonstrate that these ideal-like glasses, subjected to a dramatic thermodynamic stabilization, do exhibit the same low-temperature properties of the conventional (rejuvenated) glass. Erasing the strong structural relaxation and enthalpy reduction of pristine amber only produces a modest quantitative increase in the height of the $C_p/T^3$ peak that could be *qualitatively* ascribed to the corresponding variation of the elastic constants and the Debye coefficient. Furthermore, the best fingerprint of the universal glassy anomalies is surely the density of TLS, measured from the corresponding quasilinear contribution to the specific heat, $C_p \propto T^{1+\delta}$, since the influence of Debye-like lattice vibrations becomes less and less important below 1 K. In this respect, our experimental results are conclusive: pristine, partially-rejuvenated and fully-rejuvenated amber glasses have the same specific heat below 1 K, within experimental error.

The boson peak and the tunneling TLS are therefore robust and intrinsic properties of glasses which remain "fossilized" in 110-million-year stabilized glasses of amber, as insects or other bio-inclusions do. We also expect that amber, a hyperaged glass, will work as an extremely enlightening model glass to study many other puzzles involved in the physics of the glass state.

This work was financially supported by the Spanish MINECO (FIS2011-23488 and MAT2012-37276-C03-01 projects) and partially by the Autonomous Community of Madrid (PHAMA S2009/MAT-1756 project). T.P.C. acknowledges financial support from the Spanish Ministry of Education through the FPU grant AP2008-00030 for his PhD thesis. The authors are especially grateful to Idoia Rosales and César Menor-Salván for providing the El Soplao amber samples, to Tomás E. Gómez for his collaboration with ultrasonic experiments and to Uli Buchenau for helpful discussions and a critical reading of the manuscript. María José de la Mata is gratefully acknowledged for her technical support with TM-DSC measurements conducted at SIdI–UAM. Pilar Miranzo is thanked for her help with mass-density measurements.

---

[‡] Corresponding author: miguel.ramos@uam.es

# Supplemental Material

**1. Amber samples from El Soplao (Spain)**

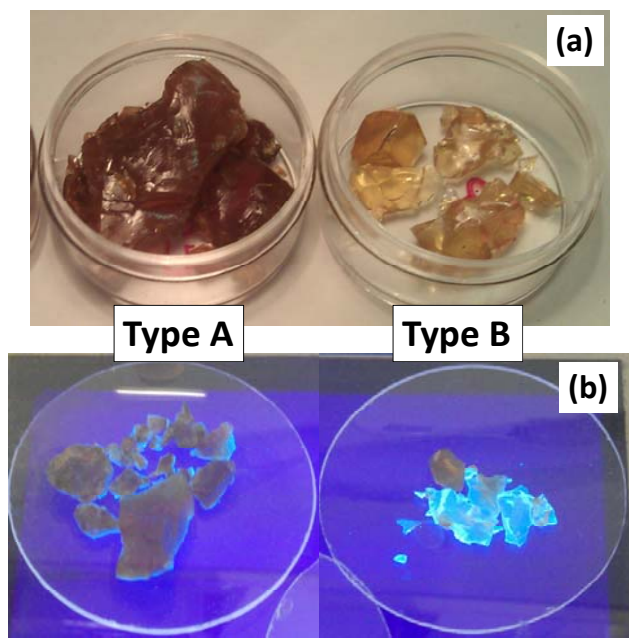

FIG. S1. The two types of Spanish amber from El Soplao deposit have their botanical origin mainly in two resin producers, namely, the extinct Cuppresaceae (type A) and Cheirolepidiaceae (type B). The differences in composition are visible to the naked eye (a). Type A amber presents a strong brownish colour under artificial light, whereas type B samples have a honey-yellow colour. (b) Under ultraviolet illumination both amber types exhibit luminescence.

New amber deposits have been discovered quite recently in the northern region of Spain, in Cantabria, within El Soplao territory [35]. These new deposits have provided a big amount of this fossil resin, yielding a large number of bio-inclusions for paleontological research [36]. Two types of amber samples have been found attending to their chemical composition, indicating the presence of two main resin producers [35]. The differences in composition between the two amber types are evident *prima facie* (Figure S1), exhibiting both of them luminescence under ultraviolet illumination.

For the studies presented in this work, aimed at looking for a possible correspondence between the stability in amber and the excess density of states at low temperatures, only the so-called "type B" [35] samples have been used, due to their better homogeneity and less amount of impurities. When needed, raw pieces of amber were cleaved or slowly cut with a diamond wheel, and then simply cleaned with distilled water.

## 2. Calorimetric characterization through TM-DSC

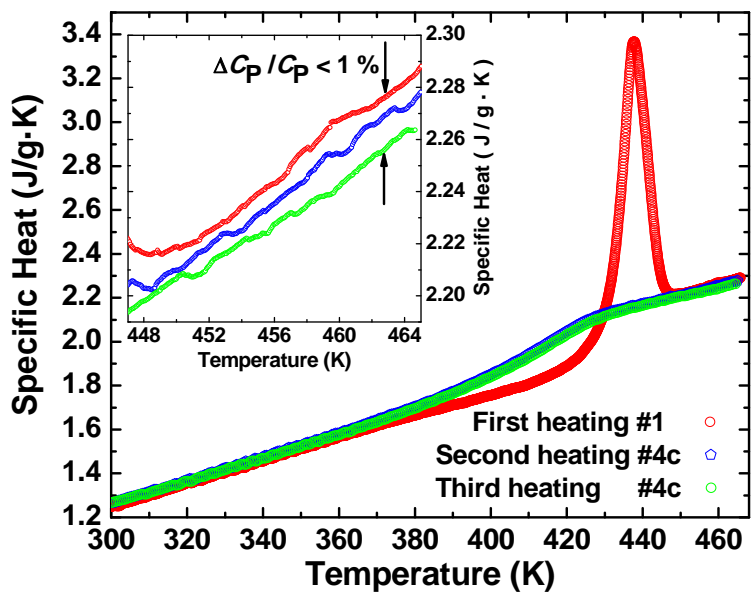

FIG. S2. Typical TM-DSC curves on a pristine amber sample (type B) corresponding to the first three consecutive heating runs, systematically performed in all cases at rates ±1 K/min and modulating signal ±0.5 K every 80 s. The first heating shows the huge endothermic signal corresponding to the strong stabilization occurred after (110 million years) *hyperaging*. The two subsequent curves correspond to the rejuvenated glass, after heating it above the glass-transition temperature. Reproducibility among them also evidences that no further repolymerization process occurs when heating the sample. The inset shows a zoom in the liquid region, where the reproducibility among different curves is determined to be ≈ 1%.

Calorimetric characterization of amber through the different stages of rejuvenation (thermodynamic destabilization) was performed using a commercial DSC Q100 from TA Instruments. The technique employed was Temperature-Modulated Differential Scanning Calorimetry (TM-DSC), which allowed us to independently determine thermodynamic and kinetic stability via reversing and nonreversing contributions to the heat capacity, as well as the total heat capacity (Fig. S2). Heating and cooling rates employed were always ±1 K/min, and modulating signals ±0.5 K every 80 s.

Obtaining accurate absolute values around 1% and high reproducibility of the specific heat was possible by paying special attention to several aspects in the experiments: (i) the standard aluminum pans (mass $m \approx 20$ mg) used where chosen to deviate less than 0.1 mg from the reference; (ii) the sample mass was maximized to be approximately 10 mg; (iii) the sample was milled down to a homogeneous grain size of 50 – 60 μm, using an agate mortar; (iv) calibration of the specific-heat curves was done by measuring standard sapphire in the very same conditions as amber, and directly comparing it to the theoretical values, what provides a temperature-dependent correction factor.

Accuracy of temperature data is estimated to be better than 0.1 K in the used TM-DSC system. Among the different methods usually employed to assess the glass transition temperature, we have determined the calorimetric (devitrification) glass transition $T_g^*$, displayed in Table I for each sample, from the maximum in the derivative of its *reversing* heat-capacity curve (first heating scan). This is the same as the inflection point of the truly (kinetic) specific-heat glass-liquid discontinuity, devoid of *non-reversing* enthalpy release also occurring around the glass-transition temperature and responsible for the (thermodynamic) fictive temperature $T_f$. The estimated error in the determination of $T_g^*$ is below ± 1K, and mainly due to the statistical uncertainty in the determination of the peak in the derivative of the *reversing* heat capacity.

Comparison between a sample annealed in the laboratory for two hours at 423 K after rejuvenation and the pristine amber shows up the much bigger effectiveness (see the huge endothermic overshoot in Fig. S3) of the natural hyperaging of amber than a laboratory time-scale annealing to stabilize the glass, therefore supporting the relevance of structural relaxation processes in glasses far below the glass transition.

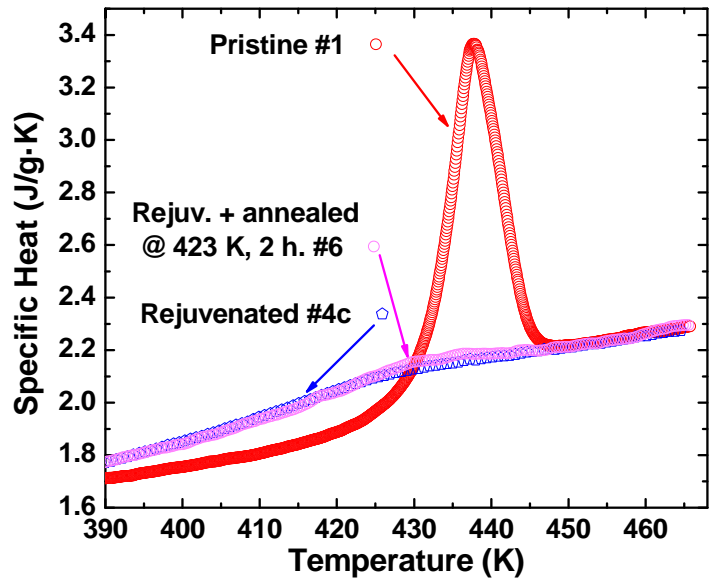

FIG. S3. Specific-heat curves corresponding to fully-rejuvenated amber of type B (sample 4c) and after annealing such a rejuvenated glass at 423 K for 2 hours (sample 6). For comparison, the pristine amber (sample 1) is also shown. A slight overshoot in the annealed rejuvenated

amber is observed, essentially negligible compared to the effect of hyperaging observed in the pristine sample.

## 3. Determination of the fictive temperature

In order to account for the thermodynamic stability of amber, i.e. how deep each glass (with a different thermal history) lies in the potential energy landscape, we have obtained the fictive temperature from the enthalpy curves. Starting with the specific-heat curves in Fig. 2a, we obtain the corresponding enthalpy curve $\Delta H(T)$ (see Fig. 2b) from direct integration $H(T) = \int_{T_0}^{T} C_p(T) dT$, $T_0$ being an arbitrary reference temperature in the liquid. In our case, $T_0 = 465$ K was taken, well above $T_g = 423$ K, and more importantly, far from the huge aging signal (centered at $T_{max} = 437$ K and width $\Delta T = 14$ K).

After the enthalpy curves are obtained, we determine the fictive temperature from the intersection between the extrapolations of the liquid (ergodic) and the glass (non-ergodic) enthalpy curves, see Fig. S4. The calculation of the extrapolations was done using experimental data far from the glass transition $T_g = 423$ K, by means of quadratic polynomial fits, given the linear behavior of the specific-heat curves well in the liquid and glass regions. Specifically, we used the temperature ranges $452 \text{ K} \leq T \leq 465 \text{ K}$ and $320 \text{ K} \leq T \leq 355 \text{ K}$ for the liquid and glass extrapolations, respectively. In Fig 2a we can observe that specific-heat curves for all shown samples (regardless of their thermal history) completely collapse for temperatures below 360 K and above 450 K.

The errors involved in the determination of $T_f$ are estimated to be $\pm 1$K. The main source of error here are the (small) statistical errors arising from the extrapolations of glass and liquid $H(T)$ curves.

It is worth noticing that a clear difference $T_g - T_f = 7$ K is observed in Table I for the rejuvenated glass, well beyond experimental error, when one might expect $T_f \equiv T_g$ for a canonical glass obtained by cooling the liquid at the same rate than the employed heating rate in the calorimetric measurement. The reason of this apparent mismatch is simply that we are sticking here to a "kinetic" definition of $T_g$ (the calorimetric devitrification temperature $T_g^*$, taken as the inflection point of the reversing heat-capacity change). With such a decoupling between *reversing* and *non-reversing* contributions to the specific heat, by means of TM-DSC, we are able to observe that the pristine, highly-stable amber glass is much more stable than the rejuvenated one, both kinetically ($\Delta T_g^* = +15$ K, i.e. the hyperaged glass only devitrifies at higher temperatures when heating) and thermodynamically ($\Delta T_f = -32$ K, implying a remarkable descent in the potential-energy landscape).

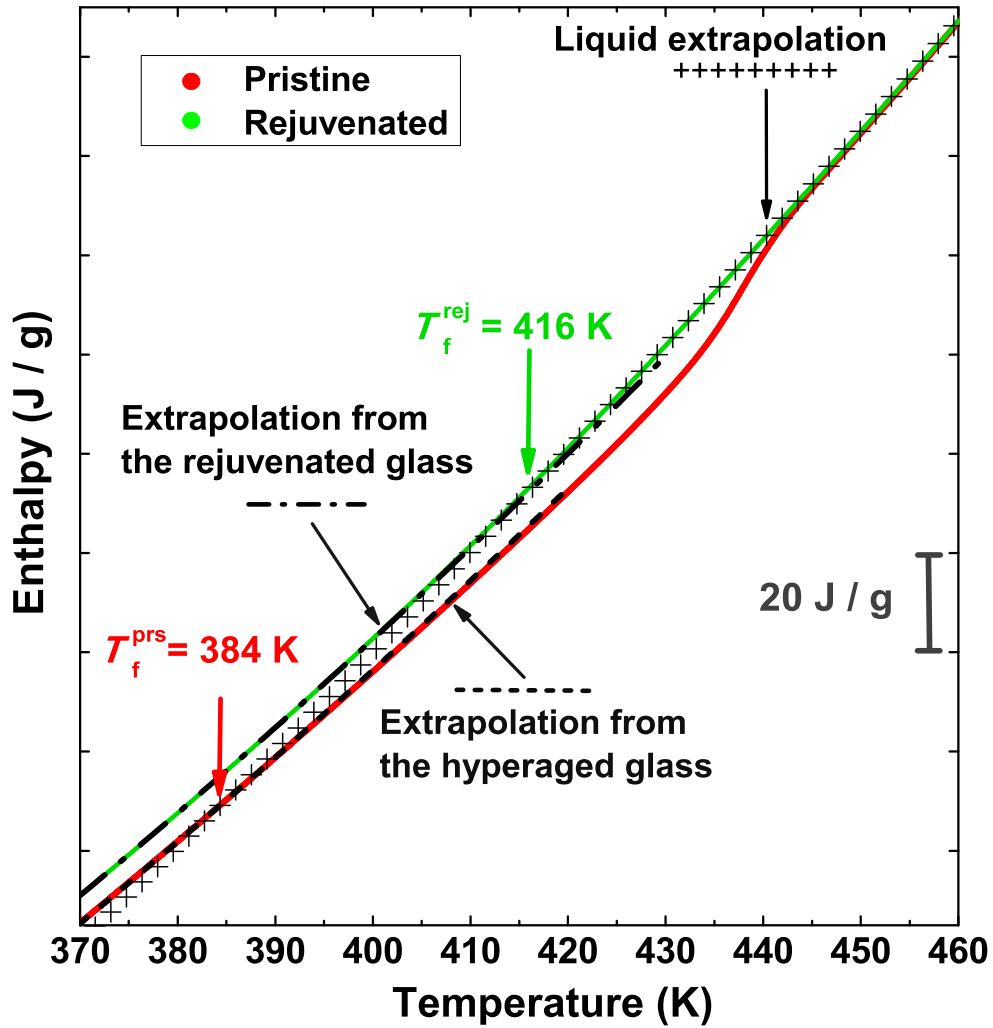

FIG. S4. Determination of the fictive temperature for the pristine ($T_f^{prs}$) and the fully-rejuvenated ($T_f^{rej}$) glass, by extrapolating the enthalpy curves from the liquid $H_{liq}(T)$ and the glass $H_{gl}(T)$ regions, far from the glass transition. The corresponding extrapolations are obtained by fitting the $H_{liq}(T)$ and $H_{gl}(T)$ curves to quadratic polynomials in the temperature ranges 452 K $\leq T \leq$ 465 K and 320 K $\leq T \leq$ 355 K, respectively.

## 4. Elasto-optic measurements

The Debye contribution to the specific heat has been independently determined for the three samples studied at low temperatures from the experimental values of the sound velocity and mass density.

The longitudinal sound velocity was measured in the temperature range 80 K ≤ T ≤ 300 K using High Resolution Brillouin Spectroscopy (HRBS), with excitation wavelength $\lambda_0$ = 514.5 nm, and extrapolated to 0 K with a least-squares fit, as shown in Fig. S4 and in Table II. Polished plan-parallel slabs of amber, less than 0.5 mm thick were placed in the same experimental setup described in Ref. [23]. Both backscattering (180º) and right-angle (90ºA) geometries were simultaneously used, the former implying a refractive-index dependent acoustic wave vector and the latter being independent of it. Given the high background signal introduced by the luminescence of the samples at these wavelenghts (see Fig. S1), the transverse sound velocities $v_T$ could not be measured by HRBS. However, the zero-temperature value $v_T(0)$ was approximately obtained by means of the generalized Cauchy equation $v_L^2(T) = A + 3v_T^2(T)$, which links longitudinal and transverse sound velocities since $A$ is found to be independent of temperature and aging in the case of polymers [37]. From ultrasonic data taken at room temperature for our amber samples, $A$=2.062 km$^2$/s$^2$ is obtained and hence $v_T(0)$ can be determined from $v_L(0)$, see Table II.

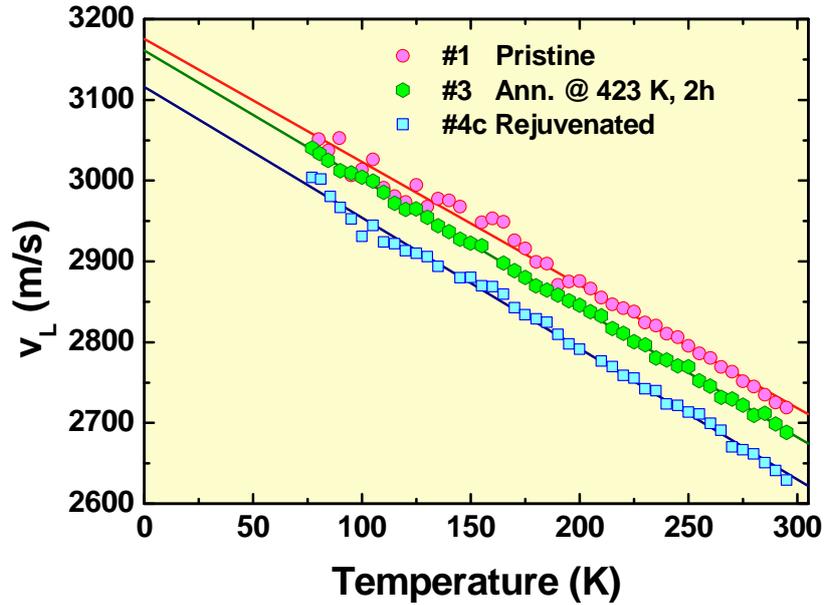

FIG. S5. Temperature dependence of the longitudinal sound velocities in Spanish type B amber with decreasing stability. Least-squares fits of the experimental data in the temperature range 80 K ≤ T ≤ 300 K are used to calculate the zero-temperature extrapolation $v_L^{prs}$ (0K) = 3175 m/s, $v_L^{ann}$ (0K) = 3160 m/s and $v_L^{rej}$ (0K) = 3115 m/s, for the pristine, annealed and fully rejuvenated samples, respectively.

The mass density of the three samples was measured at room temperature by means of the Archimedes method with a Mettler Toledo AB 265-S balance, using distilled water as a fluid. Linear extrapolation of the mass densities to zero temperature were done by recourse [23] to the Lorenz-Lorentz relation between mass density $\rho(T)$ and refractive index $n(T)$ for a transparent medium, after our HRBS measurements as a function of temperature in the range 80 K < $T$ < 300 K. The best linear fit for the experimental data was $n(T) = 1.553 - (2.619 \cdot 10^{-5}\ \text{K}^{-1}) \cdot T$, thus obtaining the $\rho(0)$ values shown in Table II. Therefore, the densification produced by the stabilization process of this amber from El Soplao is around 2%.

Then, the average Debye velocity $v_D$ was obtained, $\frac{1}{v_D^3} = \frac{1}{3}\left(\frac{1}{v_L^3} + \frac{2}{v_T^3}\right)$, and hence the calorimetric Debye coefficient in the low-temperature limit: $c_D = \frac{2\pi^2}{5}\left(\frac{k_B^4}{\hbar^3 \rho v_D^3}\right)$. As can be seen in Table II, combining mass density and sound velocity variations, the hyperaged, pristine amber has a 9% lower Debye level than the rejuvenated, canonical glass. The gradual increase of the Debye coefficient with rejuvenation goes in parallel with that exhibited by the boson-peak height, possibly suggesting that low-frequency glassy excitations and phonon-like acoustic modes are somehow mixed or hybridized at not too low temperatures. Nonetheless, it is to be remarked that even the highly-stabilized pristine amber exhibits a strong boson peak, well above the corresponding Debye level.

## 5. Low-temperature relaxation calorimetry

Low-temperature specific-heat measurements in the range 1.8K<$T$<30K were performed in a double-chamber insert, placed in a $^4$He cryostat. Measurements in the range 0.07K<$T$<3K were performed in a dilution refrigerator Oxford Instruments MX400. Both calorimetric cells consisted of a sapphire disc, on which a small calibrated thermometer (either Cernox or $RuO_2$, respectively) and a resistive chip as a Joule heater were glued diametrically opposed using cryogenic varnish GE7301. The sapphire substrate is suspended from a copper ring acting as thermally-controlled sink. The main thermal contact between them is a thin metallic wire through which heat is released. The heat capacity of the empty addenda was independently measured and subtracted from the total data points. Excellent agreement was found between experimental data from both experimental setups in the overlapping temperature range.

The low-temperature specific heat was measured in the temperature range from 0.07 K to 30 K by means of thermal relaxation calorimetry, in the experimental setup described in more detail elsewhere [23, 38]. Two slightly different methods were used depending on the relaxation time constants. For values of the order of one to a few minutes, the standard relaxation method (see Fig. S6) was employed. For higher relaxation times (typically at temperatures $T > 7$–8 K), an alternative faster relaxation method was chosen [39], in which the heating power is switched off before reaching the equilibrium. Combining the thermal charging ($\Delta T_\infty \to \kappa$) and discharging curves ($\tau$), the specific heat is obtained straightforward from $C_p = \kappa \cdot \tau$ in both cases.

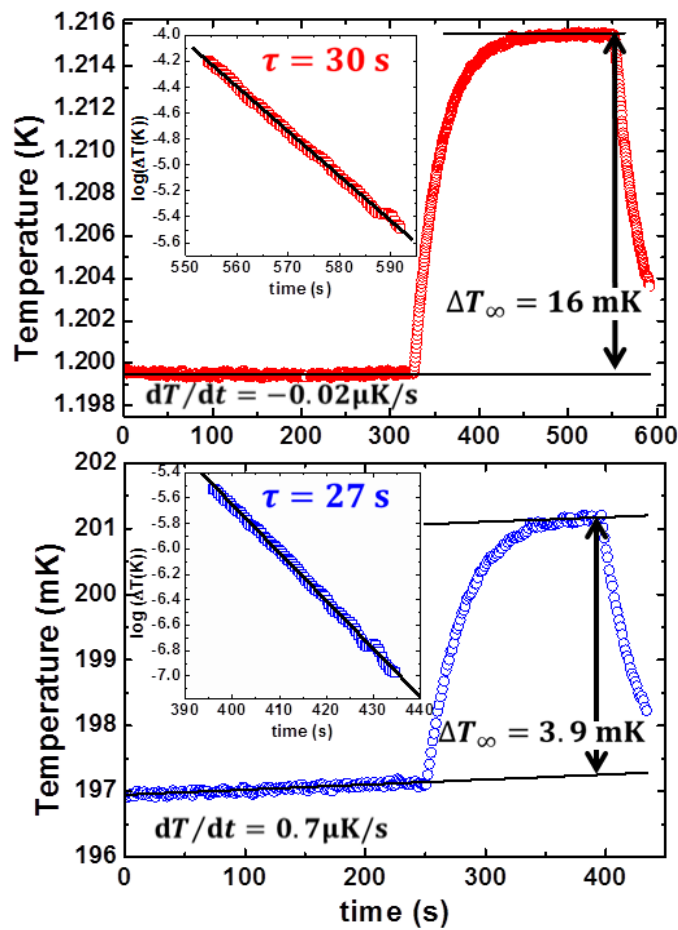

FIG. S6. Examples of raw data at 1.2 K (upper panel) and at 0.2 K (lower panel) employing the standard thermal relaxation method, used for the very low-temperature specific-heat measurements in this work. Special care in the thermal control is essential in order to accurately obtain absolute values of the specific heat, as evidenced by the low thermal drifts. Temperature jumps systematically smaller than $\Delta T/T_0 \leq 2\%$ where applied, so that the thermal conductance of the cell-to-surroundings contact could be approximated as constant in the charge-discharge process. The insets show the logarithm decrease of temperature versus time in the relaxation. The linearity of the semilogarithmic plot confirms the suitability of the method, dominated by a single relaxation time (the thermal link) via $T(t) = T_0(t) + \Delta T_\infty \cdot \exp(-t/\tau)$.

The accurate determination of the specific heat requires high thermal stability, given by drifts of the order of few hundredths to few tenths of µK/s. For the thermal conductivity of the contact to be considered constant during the whole process, temperature jumps below 2% were always applied to the calorimeter. The correct performance of the method is easily checked at every experimental point by analysing the temperature decrease versus time to be strictly linear in a semi-logarithmic scale, as shown in the insets of Fig. S6.